\documentclass[fleqn,10pt]{wlscirep}
\usepackage[utf8]{inputenc}
\usepackage[T1]{fontenc}

\title{Detecting Infrared Single Photons with Near-Unity System Detection Efficiency }

\author[1,2,$\dag$]{J. Chang}
\author[2,$\dag$]{J. W. N. Los}
\author[2]{J. O. Tenorio-Pearl}
\author[2]{N. Noordzij}
\author[2]{R. Gourgues}
\author[2]{A. Guardiani}
\author[3]{J. R. Zichi}
\author[1]{S. F. Pereira}
\author[1]{H. P. Urbach}
\author[2,3]{V. Zwiller}
\author[2]{S. N. Dorenbos}
\author[1,2]{I. Esmaeil Zadeh}
\affil[1]{Optics Research Group, ImPhys Department, Faculty of Applied Sciences, Delft University of Technology, Delft 2628 CJ, The Netherlands.}
\affil[2]{Single Quantum B.V., Delft 2628 CJ, The Netherlands.}
\affil[3]{KTH Royal Institute of Technology, Department of Applied Physics, Albanova University Centre, Roslagstullsbacken 21, 106 91 Stockholm, Sweden}
\affil[*]{\emph{Corresponding author:j.chang-1@tudelft.nl}}
\affil[$\dag$]{\emph{These authors contributed equally to this work.}}

\begin{abstract}
Single photon detectors are indispensable tools in optics, from fundamental measurements to quantum information processing. The ability of superconducting nanowire single photon detectors to detect single photons with unprecedented efficiency, short dead time and high time resolution over a large frequency range enabled major advances in quantum optics. However, combining near-unity system detection efficiency with high timing performance remains an outstanding challenge. In this work, we show novel superconducting nanowire single photon detectors fabricated on membranes with 94-99.5 ($\pm$2.07\%) system detection efficiency in the wavelength range 1280-1500 nm. The $SiO_2$/Au membrane enables broadband absorption in small SNSPDs, offering high detection efficiency in combination with high timing performance. With low noise cryogenic amplifiers operated in the same cryostat, our efficient detectors reach timing jitter in the range of 15-26 ps. We discuss the prime challenges in optical design, device fabrication as well as accurate and reliable detection efficiency measurements to achieve high performance single-photon detection. As a result, the fast developing fields of quantum information science, quantum metrology, infrared imaging and quantum networks will greatly benefit from this far-reaching quantum detection technology. 
\end{abstract}
\begin{document}

\flushbottom
\maketitle

\thispagestyle{empty}

\section*{Introduction}

Superconducting nanowire single photon detectors (SNSPDs) emerged as a key enabling technology for quantum optics experiments and photonics applications over the last two decades \cite{G2001,review1,review2}. Achieving unity system detection efficiency (SDE) with SNSPDs has been a long-standing, yet very challenging goal. For example, in quantum key distribution (QKD) systems \cite{2014MDIQKD,2017satelliteQKD1,2017satelliteQKD2}, single photon detectors with high efficiency are essential for receiving secured quantum keys over long distances. High efficiency detectors also allow closing loopholes and certify that a quantum communication scheme based on entanglement is secure \cite{entangle-qkd}. Also, for experiments requiring coincidence measurements in multiple detectors, near-unity detection efficiency is required for each channel because the multi photon count rate depends on the efficiency product of detectors involved. For example, the 12-photon coincidence count rate \cite{12photon} is about one per hour with 75\% efficiency detectors. For the same measurement, if 99.5\% efficiency detectors could be used, the coincidence count rate would be increased to one per two minutes. Similarly, in Boson sampling, single photon detectors with high SDE are required in ambitious experiments aiming for quantum supremacy\cite{boson-sampling}. Besides near-unity system efficiency, high timing performance is crucial for applications where photon arrival time is required to be precisely recorded. For example, in high-dimensional QKD\cite{HDQKD2015,HDQKD2017}, multiple bits per photon pair can be realized by encoding information in the photons' arrival times, high efficiency and time resolution are thus both required. Similarly, high timing performance is essential for improving depth resolution in light detection and ranging  \cite{lidar1,lidar2}, distinguishing signal from false counts in dark matter detection \cite{darkmatter}, enhancing the quality of quantum imaging systems \cite{imaging1,imaging2} as well as making photons with small energy difference indistinguishable for quantum erasure application \cite{quantum-eraser}. More radically, the fast-expanding quantum technologies in recent years are based on quantum states that violate local realism, as shown in\cite{bell1,bell2}, high performance SNSPDs have played important roles in experiments that successfully demonstrated loophole-free violation of Bell’s inequalities. \par

With demands from emerging applications as well as the quest to understand SNSPDs' detection limits, efforts were made in the past years to improve SDE towards unity \cite{93,NbN90,apl-photonics,95percent,98percent,98chinese}. As summarized in table \ref{tab:summary}, different material platforms were developed to achieve the highest SDE. However, achieving unity efficiency simultaneously with ultrahigh time resolution remains a challenge. Here, using a relatively thick NbTiN superconducting film and membrane cavity, we demonstrate SNSPDs with over 99\% SDE at 1350 nm (also over 98\% SDE at 1425 nm, see \href{https://}{Supplementary Material} section "List of measured devices") and above 94\% efficiency in the wavelength range  1280-1500 nm. These detectors also achieved 15-26 ps timing jitter with cryogenic amplification readout circuitry and an electrical recovery time of about 33 ns (1/e recovery time). Additionally, we clarify explicitly SNSPDs' efficiency measurement pitfalls and requirements, which will be a solid reference for single photon applications and characterization of single photon detectors.

\makeatletter
\def\thickhline{%
  \noalign{\ifnum0=`}\fi\hrule \@height \thickarrayrulewidth \futurelet
   \reserved@a\@xthickhline}
\def\@xthickhline{\ifx\reserved@a\thickhline
               \vskip\doublerulesep
               \vskip-\thickarrayrulewidth
             \fi
      \ifnum0=`{\fi}}
\makeatother
\newlength{\thickarrayrulewidth}
\setlength{\thickarrayrulewidth}{3\arrayrulewidth}

\begin{table}[htbp]
\centering

\caption{\bf Comparison of different high efficiency SNSPDs works}
\begin{tabular}{cccc}
\hline

\vtop{\hbox{\strut Material/}\hbox{\strut Temperature}} &  \vtop{\hbox{\strut SDE/}\hbox{\strut Jitter}}& Wavelength & Reference \\
\hline
WSi/120 mK & 93\%/150 ps & 1550 nm & ref.\cite{93} \\
NbN/1.8-2.1 K & 90-92\%/79 ps & 1550 nm &ref.\cite{NbN90} \\ 
NbTiN/2.5 K & 92\%/14.8 ps & 1310 nm & ref.\cite{apl-photonics}\\
MoSi/700 mK & 95\%/unknown & 1520-1550 nm & ref.\cite{95percent}\\
MoSi/700 mK & 98\%/unknown & 1550 nm & ref.\cite{98percent}\\
NbN/800 mK-2.1 K & 95-98\%/65.8-106 ps & 1530-1630 nm & ref.\cite{98chinese}\\
NbTiN/2.5 K & 94-99.5\%/15.1 ps & 1290-1500 nm & This work\\
\hline
\end{tabular}
  \label{tab:summary}
\end{table}
\section*{Optical Simulation and Device Design}

\begin{figure}[ht]
\centering
\fbox{\includegraphics[width=0.9\textwidth]{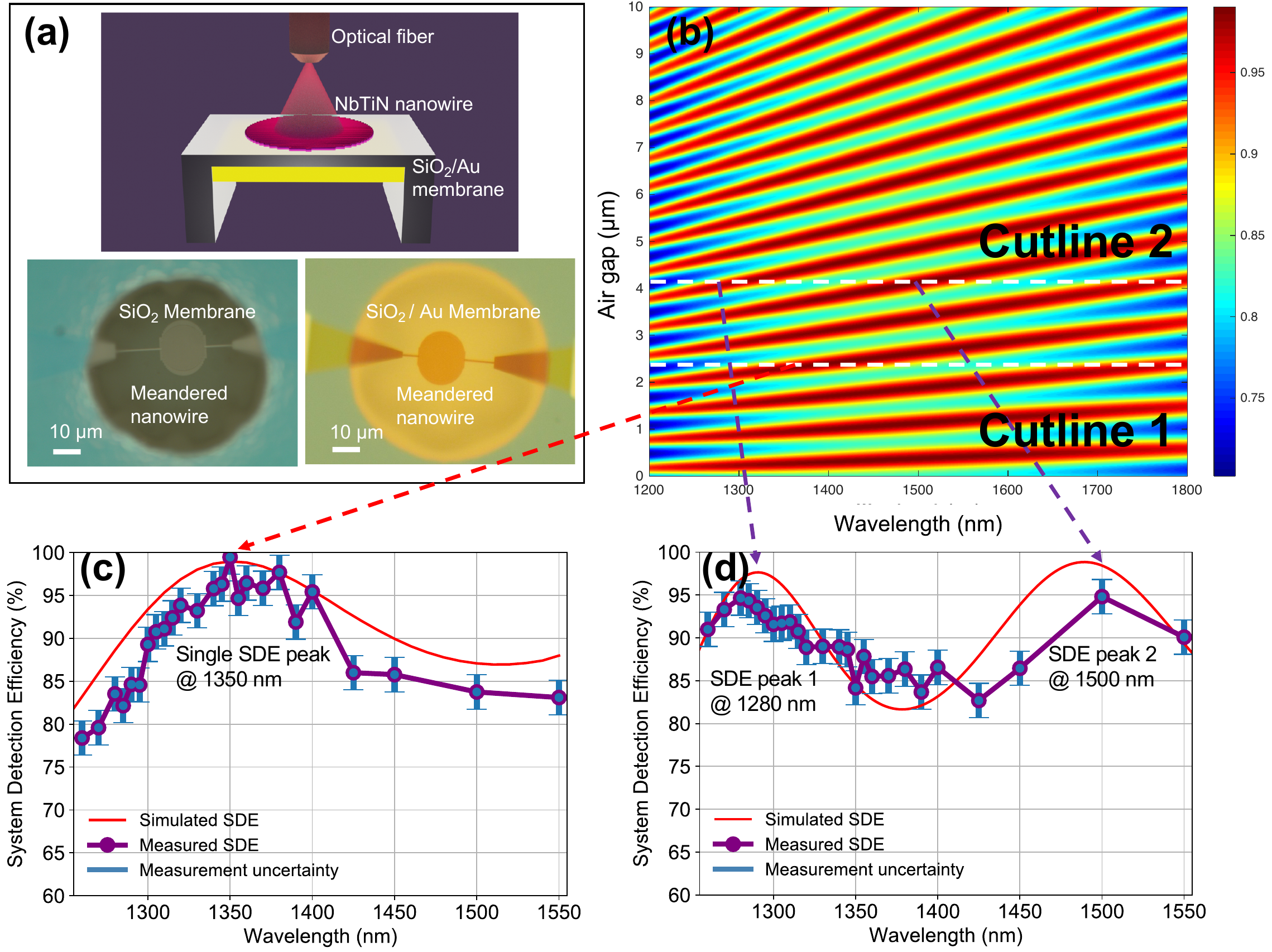}}
\caption{(a) Top panel illustrates the device structure of meandered nanowire on $SiO_2$/Au membrane and bottom panel shows optical image of the meandered nanowire on $SiO_2$ membrane before/after (left/right) Au deposition. (b) Simulated optical absorption for a device with 0-10 $\mu$m air gap. Cutline 1 shows when the air gap is around 2.2 $\mu$m, only one SDE peak occurs around 1350 nm and cutline 2 shows that with an air gap of 4.1 $\mu$m, two SDE peaks are obtained. (c) Measurement and simulation of detector \#1 with SDE over 99\% at 1350 nm. (d) Measurement and simulation of detector \#2 with dual peaks at 1280 nm and 1500nm, both exceeding 94\% SDE.}
\label{figure1}
\end{figure}

Typically, SNSPDs are meandering superconducting nanowires embedded in an optical cavity. An optimized optical cavity and meandered nanowire design are indispensable to achieve high system detection efficiency. Recent results \cite{95percent,98percent} showed that SNSPDs integrated with distributed Bragg reflectors (DBR) enable high system detection efficiency. However, since light is reflected several times in the cavity before being completely absorbed, the beam diverges and bigger detectors were made at the cost of longer recovery times (dead time in the order of few hundred nanoseconds) to have good absorption. Also, fabrication of large detectors inevitably increases the timing jitter of the detectors. More complex double-layer meanders were also reported to achieve high absorption and thus high SDE \cite{98chinese}. However, the additional meander layer increases fabrication complexity and large active areas are still needed to have strong absorption. In this work, we employ a different approach: a meandering superconducting nanowire is fabricated on a thin $SiO_2$ membrane with a Au reflector underneath. We made two different meandered nanowire designs: 50/120 nm and 70/140 nm (line-width/pitch). Both designs have a radius of 8 $\mu$m so that our device diameter is 30-70\% smaller than recent reported high SDE works \cite{98chinese,98percent}. The top panel in figure \ref{figure1} (a) illustrates the NbTiN nanowire supported by $SiO_2$/Au membrane and bottom panel shows an optical microscope image of the meandered nanowire on $SiO_2$ membrane before (left) and after (right) deposition of Au reflector. This compact optical cavity design allows us to make  smaller meandering nanowires without degrading the SDE. Also, a smaller device leads to lower kinetic inductance, translating into a faster detection signal rising edge and better timing performance. \cite{low-kenitic-inductance,electrical-model,jitter-analysis}.

\par

To achieve efficient optical fiber to detector coupling,  we used the ferule-sleeve method described in \cite{apl-photonics}. The air gap between detector and fiber plays an important role in the total optical absorption. The air gap is defined by multiple sources: (i) fabrication residuals left around the device or dust on the fiber end surface,(ii) Au contacts around the detector, (iii) potential drift of fiber core during cooling/warming, leaving a gap between detector and fiber. In this work, Finite-difference time-domain (FDTD) simulations were carried out for a systematic study of the optical absorption. As shown in figure \ref{figure1} (b), when the air gap is around 2.2 $\mu$m, only one efficiency peak can be observed along cutline 1. As a result, figure \ref{figure1} (c) shows simulation and measurement results of detector \#1 with >99\% SDE at 1350 nm. With the increase of the air gap distance, more complex absorption situations are obtained. For example, along cutline 2, dual absorption peaks are expected. We point out that an air gap doesn't always reduce absorption. With proper control of the air gap, one could achieve maximum absorption at selected wavelengths. As a direct demonstration, figure \ref{figure1} (d) shows simulation and measurments of detector \#2 with two SDE peaks: at 1280 nm and at 1500 nm. Both SDEs exceed 94\% similar to previously reported more complex cavities \cite{multispectral} and controlled design of such detectors would benefit applications where multiple wavelengths must be efficiently detected simultaneously.

\section*{Device Fabrication}
   
Based on simulation results, device fabrication was carried out as described below. Initially, a 230 nm thick $SiO_2$ layer was grown by thermal oxidation on a commercial Si wafer. On top of the $SiO_2$, a NbTiN thin film was deposited by co-sputtering of Nb and Ti in a plasma of Ar and $N_2$ as described in \cite{julien}. The meandering nanowire structure was then written by electron beam lithography with either HSQ (1st batch, negative) or ARP-62009 (2nd batch, positive) E-beam resist. After development, the nanowire pattern was transferred to the NbTiN layer by reactive ion etching with mixed gases of $SF_6$ and $O_2$. Afterwards, using deep reactive ion etching and metal evaporation, we fabricated a thin $SiO_2$ membrane with a Au mirror beneath the NbTiN nanowire as described in \href{https://}{Supplementary Material} section "Device Fabrication". Finally, a deep Silicon etch step (Bosch etching) released the detectors.   \par
 

\section*{Efficiency Measurement Setup}

\begin{figure}[!ht]
\centering
\fbox{\includegraphics[width=0.9\textwidth]{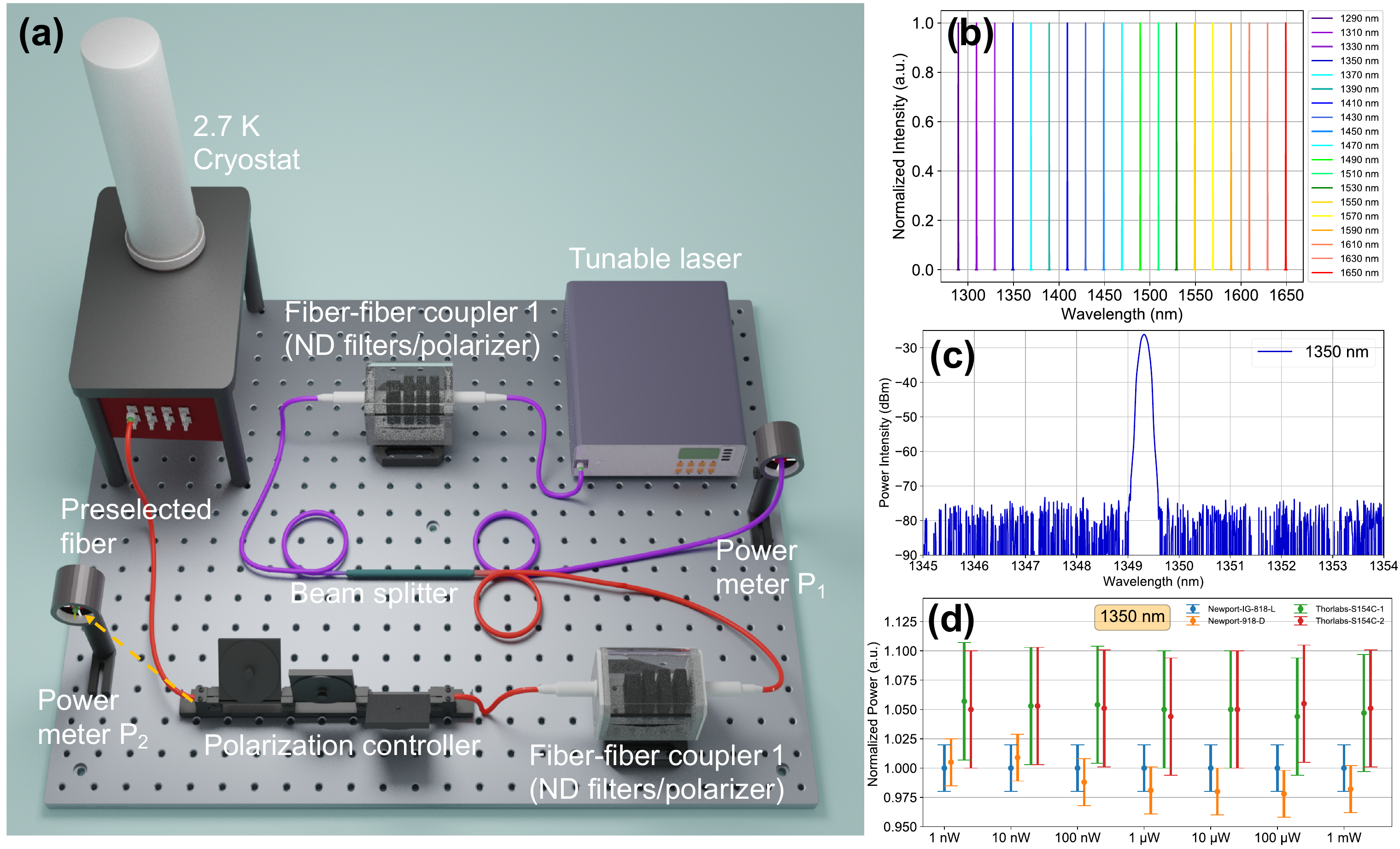}}
\caption{(a) System detection efficiency measurement setup. Emission from a tunable laser passes a bench containing neutral density (ND) filters and a polarizer, and goes through a 99-1 fiber-coupled beam splitter to split the signal towards power meter $P_1$ (99\%) and power meter $P_2$ (1\%). (b) Measured spectrum of the tunable laser at 1290-1650 nm. (c) Measured laser spectrum at 1350 nm shows a laser linewidth of < 1 nm. (d) Four different optical power meters' readings at 1350 nm with different error bars. All readings are normalized to the Newport IG-818-L power meter.}
\label{figure2}
\end{figure}

Prior to any measurement, the laser was turned on for > 1 hour for power stabilization. Every optical component including fibers were fixed to avoid influence from mechanical vibration and air turbulence. We used the following two-step procedure to carry out system efficiency measurements: (i) Building an accurate laser attenuator. Initially, the continuous-wave (CW) laser beam passes through the first fiber-to-fiber coupler (FBC-1550-FC, containing a polarizer) followed by a fiber-coupled beam splitter (with a splitting ratio of 99-1). The high power branch is recorded by an optical power meter $P_1$ while the low power branch is directed towards the second fiber-to-fiber coupler (containing neutral density filters and a polarizer) and a polarization controller. Similarly, the power after the polarization controller is recorded with an optical power meter $P_2$. By adjusting polarizers in both fiber-to-fiber couplers and choosing the proper neutral density filters, we set the power ratio $P_1$/$P_2$ to the desired values (50-60 dB). After the attenuation ratio was set, all components were kept fixed. (ii) Controlling precisely the input photon flux. We lowered the input power by adding extra neutral density filters before passing through the first polarizer to lower $P_1$ to 1-10 nW. We also rechecked the attenuation ratio multiple times before and after the measurements to assure nothing has been changed in the setup (see \href{https://}{Supplementary Material} section "Efficiency measurement stability"). After this two-step procedure, different input photon fluxes can be set up, for example 10 nW with 50 dB attenuation corresponds to 679k photons per second at 1350 nm. It must be noted that, to avoid fiber-to-fiber coupling losses when connecting to the detection system, we preselected the fibers which have the best coupling match to the fibers inside our cryostat. Finally, the fiber at the output of the polarization controller was connected to the preselected fiber and then guide light into the system for measurements. \par 
Our SDE was calculated as $\eta_{SDE}$=(1-$R_{rfl}$)$\cdot$($N_{count}$/$N_{total}$), where $N_{count}$ is the total registered count rate by our system and $N_{total}$ is the total input photon number. $R_{rfl}$ is added to avoid overestimation of the SDE and it represents the simulated and measured fiber-air interface reflection (see \href{https://}{Supplementary Material} section "Fiber end-face reflection"). Since the total input photon flux was calculated by $N_{total}$ = P $\cdot$ $\lambda$/(hc), where P is the measured optical power, h is plank constant, c is the speed of light in vacuum and $\lambda$ is the used wavelength, we carefully evaluate our laser spectrum with an optical spectrum analyzer. In figure \ref{figure2} (b), we show the measured spectrum of the tunable laser at variable wavelenghts, from 1290 to 1650 nm. As a result, figure \ref{figure2} (c) demonstrates that the laser has a linewidth of < 1 nm. The slight shift of the measured wavelength with the set value is mainly due to the optical spectrum analyzer's calibration off, which has negligible influence on our SDE measurement. For input power measurement, accurate optical power meters are necessary. As shown in \ref{figure2} (d), four different optical power meters' readings at 1350 nm are presented with their uncertainties. All readings are normalized to the Newport IG-818-L (used for our measurements) since  it has 2\% accuracy from 20 pW until 10 mW and a good linearity uncertainty of 0.5\%.

\section*{Detection performance and Discussion}

\begin{figure}[!ht]
\centering
\fbox{\includegraphics[width=0.8\textwidth]{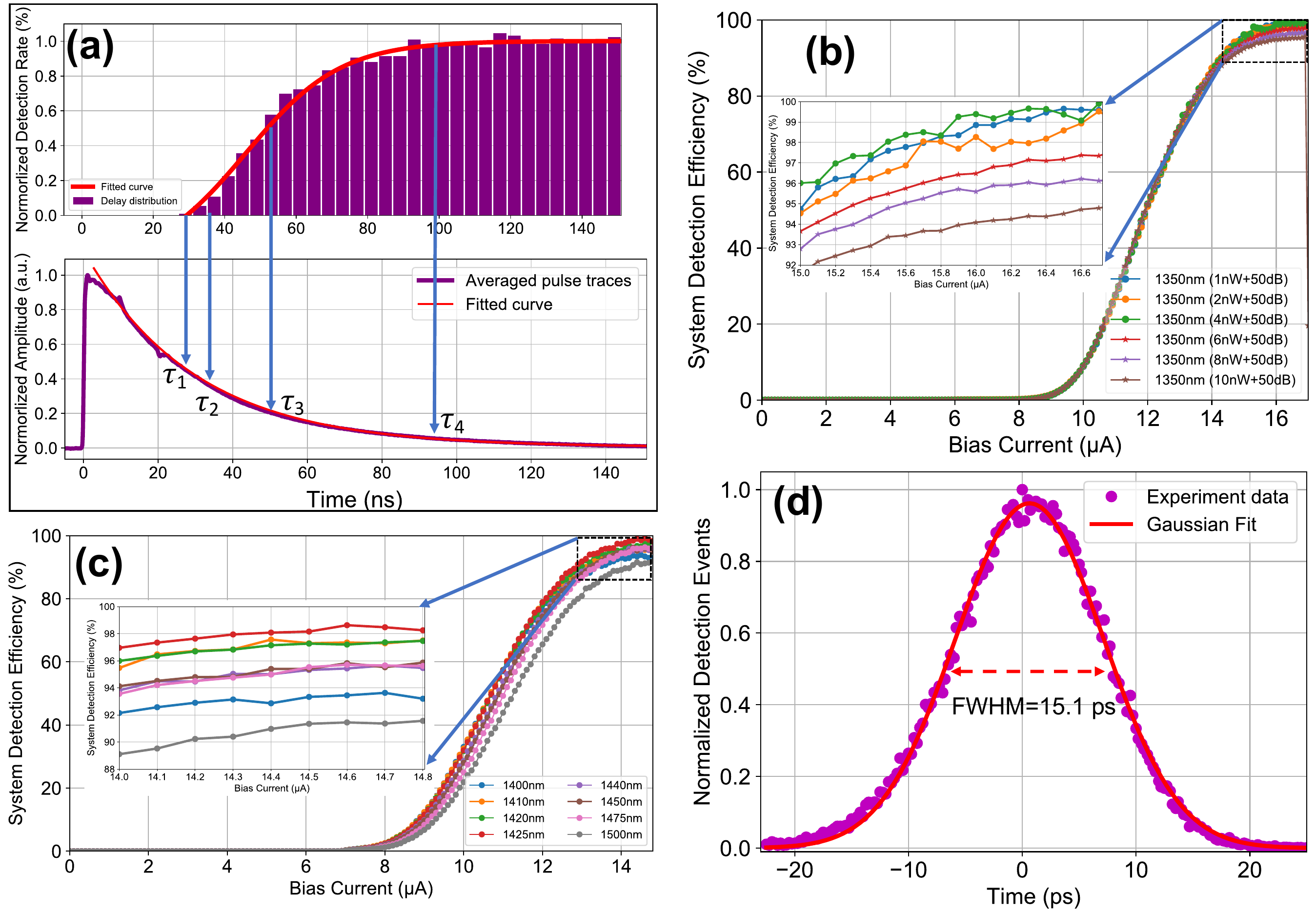}}
\caption{(a) The top inset shows the auto-correlation measurement of a detector indicating SDE recovery dynamics. The bottom inset shows an averaged pulse trace from the same detector. (b) SDE measurements of detector \#1 at 1350 nm with different input photon fluxes. When input photon flux is below 4 nW plus 50 dB attenuation, the SDE of detector \#2 reached >99\%, which is also shown in the inserted picture. (c) SDE measurements of detector \#3 at different wavelengths. The maximum SDE of detector \#3 reached >98\% at 1425 nm. (d) Jitter measurement from a detector with > 91\% SDE with cryogenic amplifier. A Gaussian fit gives a FWHM jitter of 15.1 ps.}
\label{eff}
\end{figure}

Prior to SDE measurements, we studied the relationship between SNSPDs detection reset kinetics and their efficiency recovery. We performed auto-correlation measurements between two subsequent detection events with the detectors illuminated by a CW laser similar to \cite{auto-corre}. After collecting more than 50 thousand events, as shown on the top panel of figure \ref{eff} (a), we built a delay time histogram. Together with the SNSPD detection pulse shown at the bottom panel of \ref{eff} (a), we can see that within the first 25 ns, no subsequent pulse can be detected thus $\tau_1$=25 ns can be defined as the minimum separation dead time. For most SNSPD related studies, 1/e dead time (time for pulse to decay from peak amplitude to 1/e of the amplitude) is often used to describe the device recovery property. It can be seen that our device's 1/e dead time is $\tau_2$=33 ns. Furthermore, $\tau_3$=51 ns represents the time when the detector recovers 50\% of the maximum efficiency, also known as -3dB efficiency dead time and $\tau_4$=97 ns stands for full efficiency recovery time. These measurements indicate that input photon flux can influence  SDE because if the photon flux is too high, photons arriving within the dead times of the detector can not be registered at the detectors' maximum efficiency, thus optimal input photon flux is necessary to achieve maximum detection efficiency. It must be noted that much higher photon fluxes can be achieved (with no loss of efficiency) if the source is pulsed (photons arriving with regular timings in between) \cite{apl-photonics}. \par 
We characterized 40 detectors in two separate fabrication rounds. As shown in figure \ref{eff} (b), detector \#1 was the best detector from the first fabrication batch and was tested at 1350 nm with different input photon flux. Initially, the photon flux was set to 10 nW plus 50 dB attenuation ($\sim$ 679,000 photons/s), and device \#1 showed SDE of 94-95\%. With an input photon flux below 4 nW plus 50 dB attenuation ($\sim$ 271,600 photons/s), device \#1 achieved saturated SDE of >99\%. Similarly, detector \#3 from the second fabrication batch was tested at different wavelengths and showed >98\% SDE at 1425 nm as shown in figure \ref{eff} (c). For more examples, see \href{https://}{Supplementary Material} section "List of measured devices". \par   
Besides high SDE, time resolution is another crucial advantage of SNSPDs compared with other single photon detectors. The Instrument Response Function (IRF) of our detectors were characterized with a ps-pulse laser (1064 nm) and a fast oscilloscope (4 GHz bandwidth, 40 GHz sampling rate) as described in \cite{apl-photonics}. As shown in \ref{eff} (d), with a low-noise cryogenic amplifier mounted at 40 K stage in the same cryostat , the IRF of device \#15 shows a Gaussian shape histogram, after fitting we obtain 15.1$\pm$0.05 ps (full width at half maximum, FWHM) timing jitter. This detector was measured to have more than 91\% SDE and the ultra low timing jitter was mainly achieved by fabricating relatively small detectors, which results in lower kinetic inductance and thus better jitter \cite{electrical-model,jitter-analysis}. For more statistics of jitter measurement, see \href{https://}{Supplementary Material} section "Overview of tested detectors". In short we achieved 15.1-26 ps jitter with cryogenic amplifierd and 29-39 ps jitter with room-temperature amplifiers readout circuitry.


\section*{Conclusion}
In conclusion, we demonstrated NbTiN based SNSPDs operated at 2.7K with over 94\% system efficiency in the wavelength range 1280-1500 nm with the highest system efficiency of 99.5$\pm$2.07\%. At the same time, our detectors showed timing jitter in the range 15.1-26 ps with cryogenic amplifiers. The ultrahigh efficiencies were achieved using the following methods: (i)  optimized thick NbTiN superconducting film with saturated internal efficiency, (ii) optimized broadband membrane cavity coupled to small detectors, and (iii) accurate system efficiency measurements with a narrow linewidth tunable laser to precisely locate the high-efficiency peaks. Compared with previous reported high-efficiency SNSPDs \cite{93,95percent,98percent}, our work presents a platform with higher operation temperature (2.7 K, compatible with compact closed-cycle cryostats), short recovery time, and high timing resolution. At the same time, the system efficiency performance of our devices are inpar with recently reported NbN based SNSPDs \cite{98chinese} but using simpler fabrication (single-layer meander), higher operation temperature (no need for mK cooler), and better timing resolution. Our detectors can be further developed by considering the following aspects: (i) multiplexing detector and control individual pixels by cryo-CMOS electronics to realize imaging at the single photon level (ii) extending the detection spectrum in the mid-infrared by tailoring and optimizing NbTiN films and (iii) improving working temperature of the detectors with novel superconducting materials. 
\section*{Methods}
To achieve accurate SNSPD efficiency measurements, we addressed the following aspects separately.

\section*{Optical Simulation}
To simulate the absorption of the optical stack, we used the commercially available FDTD Solutions software from Lumerical. The SNSPD was modeled as the cross-section of a single nanowire in an optical cavity. From top to bottom, the simulated stack structure was: optical fiber layer ($SiO_2$), airgap, NbTiN meander, 1/4 $\lambda$ $SiO_2$ layer, and a 200 nm thick Au mirror. For more details about simulation settings and parameters, see \href{https://}{Supplementary Material} section "Optical Simulation".

\subsection*{Laser Source}
A tunable laser (JGR-TLS5) with attenuation was employed as a quasi single photon source. The laser covers the range 1260-1650 nm with a step size of 0.1 nm (FWHM). For more details on the laser, see \href{https://}{Supplementary Material} section "Tunable laser source". Compared with previous work \cite{apl-photonics} which used photodiodes operated at a single wavelength, the tunable laser has two major advantages: First, its narrow spectrum (<1 nm) at a tuned frequency allows for precise measurements; Second, a laser with tunable wavelength enabled mapping system efficiency at different wavelengths. In this way we precisely determine peak efficiency and built the spectral response.

\subsection*{Optical Power Meters}
A semiconductor-based optical power meter was the key reference for efficiency measurements. In this work, we used two different types of power meters: Thorlabs S154C (NIST traceable, $\pm$5\% uncertainty), Newport 818-IG-L (NIST traceable, $\pm$2\% uncertainty). However, for the measurement, one should not only take the power meter accuracy into consideration, but also consider sensor linearity, spectral range, power range, stability and all other related parameters. For details, see \href{https://}{Supplementary Material} section "Optical power meters".


\subsection*{Measurement Uncertainty Calculation}
For system detection efficiency measurement uncertainty, we considered all possible uncertainties in our experiments and calculated the total measurement uncertainty with the root-mean-square (RMS) of the sum of the squared errors. The uncertainties in our measurements include the power meter measurement uncertainty (2\%) and linearity uncertainty (0.5\%), laser stability uncertainty (<0.1\%) and optical attenuator uncertainty (<0.2\%). For detailed measurement uncertainty calculations, see \href{https://}{Supplementary Material} section "Efficiency measurement stability" and "Measurement uncertainty". In short, our efficiency measurement has a total uncertainty of $\pm$2.07\% (RMS).

\subsection*{Fiber end-face reflection}
When measuring laser power within the fiber with a power meter, the fiber end-facet was not in direct contact with the power meter's sensor. The existing fiber-to-air interface leads to a reflection up to a few percent back towards the light source. \cite{apl-photonics}. On the other hand, Physical Contact polished fiber to fiber connections have negligible back reflections (typically -30 to -40 dB). For all our efficiency measurements, we removed this back reflection contribution by multiplying a correction factor of (1-$R_{rfl}$). To determine the accurate value of $R_{rfl}$, see \href{https://}{Supplementary Material} section "Fiber end-face reflection".

\subsection*{Polarization Degree}
Since our detectors were patterned along meandering shapes, light absorption can be significantly different based on the light polarization direction along the meander's direction \cite{pl}. Thus it is important to have a linearly polarized input light and fully control the polarization. In \href{https://}{Supplementary Material} section "Polarization degree and control" we show detailed measurement of degree of linearity and polarization control.

\bibliography{main}

\section*{Acknowledgements} 
J.C. acknowledges China Scholarships Council (CSC), No.201603170247. I.E.Z., V.Z., and Single Quantum B.V. acknowledge the supports from the ATTRACT project funded by the EC under Grant Agreement 777222. I.E.Z. acknowledges the support of Nederlandse Organisatie voor Wetenschappelijk Onderzoek (NWO), LIFT-HTSM (Project 680-91-202). R.B.M.G. acknowledges support by the European Commission via the Marie-Sklodowska Curie action Phonsi (H2020-MSCA-ITN-642656). V.Z. acknowledges funding from the Knut and Alice Wallenberg Foundation Grant “Quantum Sensors”, and support from the Swedish Research Council (VR) through the VR Grant for International Recruitment of Leading Researchers (Ref 2013-7152) and Research Environment Grant (Ref 2016-06122).



\section*{Competing interests} 
The authors declare no conflicts of interest.


\end{document}